\documentclass{article}
\usepackage{ijcai26}

\usepackage{times}
\usepackage{soul}
\usepackage{url}
\usepackage[hidelinks]{hyperref}
\usepackage[utf8]{inputenc}
\usepackage[small]{caption}
\usepackage{graphicx}
\usepackage{amsmath}
\usepackage{amsthm}
\usepackage{amsfonts}
\usepackage{booktabs}
\usepackage{algorithm}
\usepackage{algorithmic}
\usepackage[switch]{lineno}
\usepackage{subfig}
\usepackage{bbm}
\usepackage{float}
\usepackage{xcolor}
\usepackage{enumitem}

\newtheorem{lemma}{Lemma}

\title{SymVD: Symmetric Vision–Language–Action Distillation
for Robot Manipulation 
}

\author{
Hyewon Choi$^{1, *}$
\and
Donggyu Kim$^{2, *}$
\And
Soojean Han$^{3, *}$\\
\affiliations
$^*$KAIST\\
\emails
chw110825@kaist.ac.kr,
kimd58050@kaist.ac.kr,
soojean@kaist.ac.kr
}
\begin{document}

\maketitle


\begin{abstract}
    While pretrained Vision-Language-Action (VLA) models offer broad generalization capabilities in robotic manipulation tasks, adapting them to real-world environments or handling task shifts often requires substantial additional data and retraining. 
    To address this, we propose Symmetric VLA Distillation (SymVD), a distillation framework that transfers knowledge from a large VLA teacher to a compact student policy by explicitly exploiting geometric symmetries in manipulation tasks, such as rotational and reflectional invariance. 
    SymVD employs an equivariant actor-critic architecture and trains the student using a symmetry-aware objective that aligns with teacher actions under group-invariant properties. We demonstrate that by enforcing the policy to respect equivariance, SymVD reduces redundant exploration across configurations related by group transformations and improves sample efficiency during distillation. 
    To further stabilize and improve distillation, SymVD introduces an adaptive weighting scheme that dynamically balances the distillation objective and reinforcement learning updates based on training progress, enabling robust transfer even when the teacher signal is imperfect or misaligned.
    Experimental results on robotic manipulation tasks demonstrate that SymVD consistently improves over standard distillation and also outperforms SAC in terms of sample efficiency and generalization to previously unseen symmetric transformations of the environment.
\end{abstract}

\section{Introduction}
The emergence of Robot Foundation Models (RFMs), including Vision–Language–Action (VLA) models, has reshaped modern robotics by enabling robots to handle a broad spectrum of 
tasks with improved generalization~\cite{firoozi2025foundation,kim2024openvla,team2024octo}. 
In particular, for manipulation tasks, these models learn to interpret visual scenes and follow natural-language instructions through large-scale pre-training on diverse datasets, demonstrating strong generalization to unseen objects and novel commands~\cite{yang2025transferring,liang2025vla}. However, this generalization often fails to translate to real world deployment, where even minor changes in domain or task specification require retraining on substantial amounts of data~\cite{brohan2023rt2}.
This lack of flexibility contrasts with the demands of real-world robotic systems, which require compact, adaptable policies that are easily updated~\cite{xiao2025}.

\begin{figure}[!t]
    \centering
    \includegraphics[width=\columnwidth]{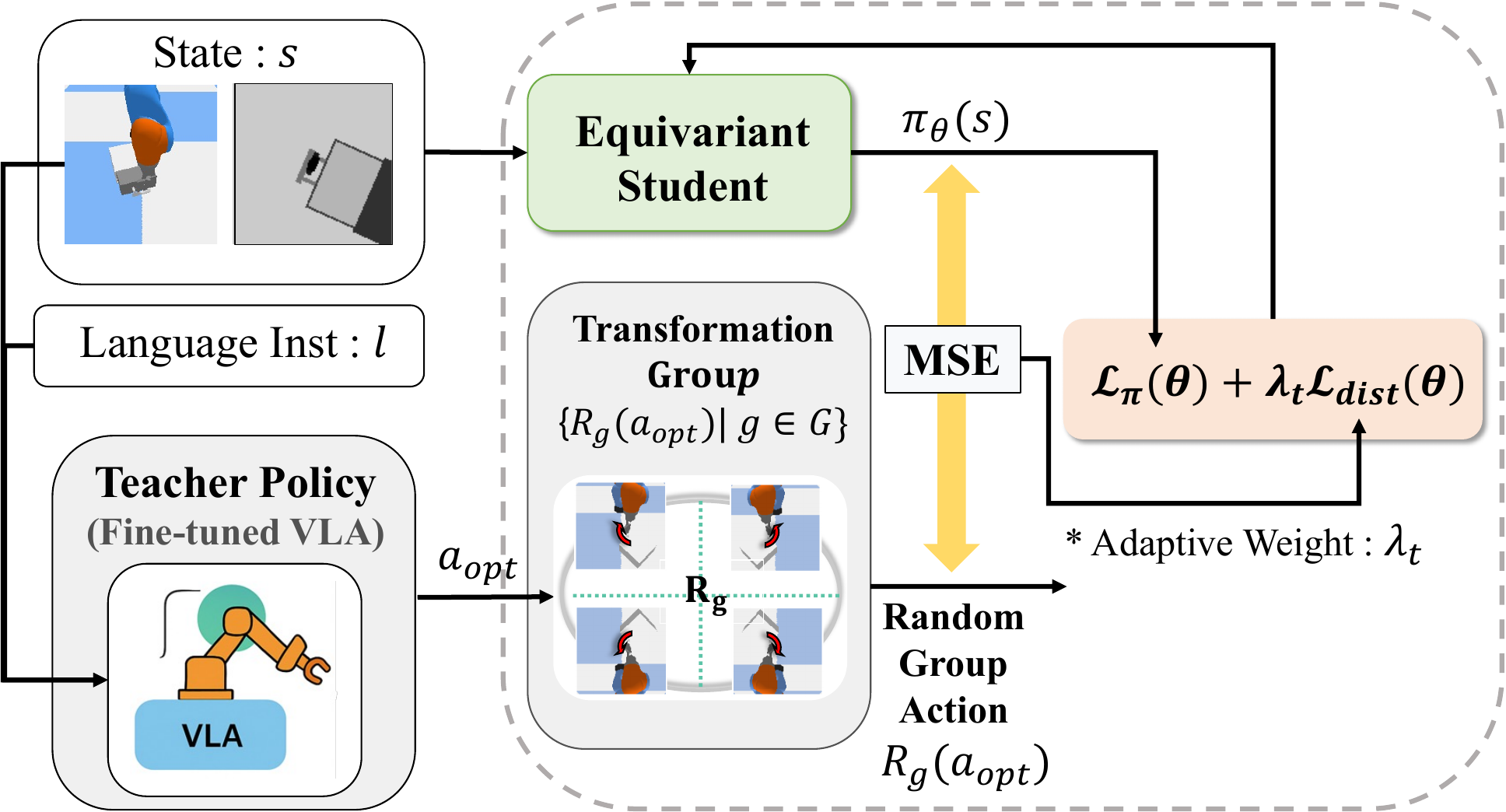}
    \caption{An overview of SymVD. A fine-tuned VLA teacher provides partially optimized actions, which are expanded into a transformation group orbit. An equivariant student is then trained by minimizing a MSE distillation loss to match these symmetry-consistent targets, combined with the RL objective. 
}
    \label{fig:results}
\end{figure}

A promising direction to address this limitation is policy distillation, where a reinforcement learning agent (student) is trained to imitate partially optimized actions from a large foundation model (teacher), thereby distilling the teacher's knowledge into a smaller, more efficient policy~\cite{rusu2015policy,gu2024minillm,agarwal2024onpolicy}. Although this teacher-student framework can speed up the initial stage of learning by mimicking expert behavior~\cite{rusu2015policy}, distilled agents often still rely on heavy exploration to perform well under domain or task shifts, resulting in sample inefficiency~\cite{xiao2025}. One direction to mitigate this issue is equivariant learning, which exploits structural regularities,
such as geometric symmetries, so that the student can share behavior across configurations that are equivalent under transformations~\cite{mondal2020group,mondal2022eqr}. While structural regularities have been widely studied in the context of equivariant learning, their role in distilling generalist policies into symmetry-aware agents remains underexplored. 

Building on these observation, we propose the Symmetric VLA Distillation (SymVD) model, a framework that unifies the broad generalization capability of pretrained VLAs with the data efficiency of symmetry-aware RL agents. The central idea is to distill a teacher's policy into a student whose neural architecture is designed to respect the geometric symmetries of the underlying task through group-equivariant design. To this end, SymVD extends \texttt{E2CNN}~\cite{weiler2019general} to policy learning for high-dimensional continuous control. It then trains the student via a supervised loss to align with the teacher while preserving symmetry constraints, and employs an adaptive weighting scheme to dynamically balance the distillation objective and RL updates for stable and robust training. By constraining the policy search space, SymVD reduces the amount of trial-and-error exploration and improves transfer to unseen configurations related by symmetry transformation.

Our main contributions are as follows:
\begin{itemize}
\item We introduce SymVD, a framework that integrates VLA-based policy distillation with an equivariant actor–critic architecture to enhance sample efficiency in robotic manipulation.
\item We develop a symmetry-aware distillation objective that aligns teacher actions with the representational constraints of an equivariant student, enabling more effective knowledge transfer than conventional distillation.
\item We propose an adaptive distillation weighting mechanism that dynamically balances distillation and reinforcement learning updates, improving training stability and robustness when the teacher signal is imperfect.
\end{itemize}

The remainder of the paper is organized as follows.
In Section~\ref{sec:method}, we first formalize robotic manipulation as a group-invariant MDP and define the group actions on states and actions. We then present the SymVD framework, detailing the equivariant actor-critic architecture and the proposed group-averaged distillation objective with adaptive weighting.
Section~\ref{sec:experiments} evaluates SymVD on BulletArm manipulation tasks using both VLA and lightweight teachers, and provides an analysis of the adaptive weighting mechanism.
Finally, we conclude the paper in Section~\ref{sec:conclusion}.

\section{Related Work}
\subsection{Policy Distillation for Generalist Policies}
Policy distillation has been widely studied as an effective mechanism for transferring knowledge from an expert policy to a compact student model~\cite{rusu2015policy,Levine2014}. Recent advances~\cite{torne2024,zhou2024archer,mark2024policy} have extended policy distillation to the fine-tuning and adaptation of large-scale foundation models. While foundation models offer strong generalization, their high computational overhead and poor domain adaptability often preclude direct real-world deployment. Reinforcement learning (RL) complements this by enabling experience robot behavior acquisition in safe simulation~\cite{torne2024,xu2024rldg,kalashnikov2018scalable}, and naturally incorporating task-specific reward/value signals without relying on human feedback~\cite{gupta2019,nakamoto2024}. There is also growing interest in distilling a generalist's knowledge into lightweight RL policies, with recent frameworks~\cite{julg2025refined,brohan2023rt2,xiao2025} demonstrating faster exploration and improved deployability. However, RL-based distillation approaches still remain highly sensitive to the quality of the teacher policy. Suboptimal or biased demonstrations from the teacher can lead to degraded student performance, a phenomenon highlighting the need for more resilient distillation objectives~\cite{julg2025refined,agarwal2024onpolicy}.

\subsection{Equivariant Learning in Robotics}
Exploiting structural regularities such as geometric symmetries has been shown to significantly improve generalization and sample efficiency in robotic learning.
In visual perception for robotic manipulation, symmetry has been extensively exploited to improve generalization, including SO(2)-equivariant representations for top-down visual inputs, SE(3)-invariant grasp evaluation, and harmonic--based models for continuous grasp directions~\cite{huang2022equivariant,huang2022edgegrasp,hu2024orbitgrasp}.
Numerous robotic studies have incorporated group symmetries into RL agents by designing invariant or equivariant policies, enabling them to share experience across states and actions equivalent under symmetry transformations~\cite{mondal2020group,mondal2022eqr}. However, imposing structural symmetry through constrained observation representations narrows the exploration space, making policy learning difficult to converge without guidance from accurate planner-based experts~\cite{wang2022mathrm}. Prior symmetry-aware approaches typically focus on learning policies from scratch and the distillation of knowledge from generalists
has remained relatively underexplored. Our study bridges this gap by combining VLA-based policy distillation with equivariant policy architectures, enabling efficient and symmetry-consistent knowledge transfer from generalist teacher models.

\section{Method}
\label{sec:method}
\subsection{Problem Formulation}
\label{subsec:problem_formulation}

We formulate the robotic manipulation task as a Markov Decision Process (MDP) augmented with symmetry constraints. Consider a MDP tuple $\mathcal{M} = (\mathcal{S}, \mathcal{A}, r, p, \gamma, G)$, where $\mathcal{S}$ and $\mathcal{A}$ denote the state and action spaces, $r(s,a)$ is the reward function, $p(s'|s,a)$ is the transition kernel, $\gamma$ is the discount factor, and $G$ is a symmetry group acting on the system.

Let $L_g : \mathcal{S} \rightarrow \mathcal{S}$ and $R_g : \mathcal{A} \rightarrow \mathcal{A}$ be the group transformations associated with an element $g \in G$. The MDP is said to be \textit{group-invariant} if the dynamics and rewards are preserved under these transformations for all $g \in G$, $s \in \mathcal{S}$, and $a \in \mathcal{A}$:
\begin{equation}
\label{eq:group-invariant}
\begin{aligned}
    r(s,a) &= r(L_g(s), R_g(a)), \\
    p(s' \mid s,a) &= p(L_g(s') \mid L_g(s), R_g(a)).
\end{aligned}
\end{equation}
This formulation implies that a state-action pair $(s,a)$ is physically equivalent to its transformed counterpart $(L_g(s), R_g(a))$. By leveraging these geometric symmetries, the agent can generalize experienced transitions across all symmetric states, significantly improving sample efficiency.

In robotic manipulation, we consider a state space $\mathcal{S}$ represented as a 2D image over $\mathbb{R}^2$, and a continuous action space $\mathcal{A} \subset \mathbb{R}^d$. While our formulation applies to general symmetry groups, we focus on planar rotational symmetries in this work, dictated by the $\mathrm{SO}(2)$-equivariant design of the student policy architecture. An element of $\mathrm{SO}(2)$ corresponds to a rotation matrix $R(\theta)$ for $\theta \in [0, 2\pi)$.
%
Our objective is to learn a policy $\pi_\xi$ that maximizes expected returns while adhering to these geometric constraints. Furthermore, we aim to distill symmetry-consistent behaviors from a pretrained VLA policy $\pi_{\text{VLA}}$ into our equivariant student agent.

\subsection{Equivariant Agent Design}
To effectively exploit the workspace symmetries defined in Section~\ref{subsec:problem_formulation}, we adopt the Equivariant Actor-Critic (Equi-SAC) framework~\cite{wang2022mathrm}. For computational tractability, we replace the continuous rotation group $\mathrm{SO}(2)$ with a finite cyclic approximation $C_k$. In our experiments we use $C_4$ (rotations by multiples of $90^\circ$), but the same construction extends straightforwardly to finer discretizations such as $C_8$ (multiples of $45^\circ$) or, more generally, any $C_k$.

We decompose the action space into equivariant and invariant subspaces,
$\mathcal{A} = \mathcal{A}_{\mathrm{eq}} \times \mathcal{A}_{\mathrm{inv}}$,
and represent an action as a tuple
$a = [\mathbf{a}_{xy}^\top, \mathbf{a}_{\mathrm{aux}}^\top]^\top$.
The planar translation $\mathbf{a}_{xy} = (a_x, a_y) \in \mathcal{A}_{\mathrm{eq}}$ transforms equivariantly under workspace rotations, whereas auxiliary dimensions such as gripper aperture, vertical displacement, and end-effector rotation are treated as group-invariant, denoted by
$\mathbf{a}_{\mathrm{aux}} = (a_z, a_\theta, a_\lambda) \in \mathcal{A}_{\mathrm{inv}}$.
($d=5$)
Under the discrete rotation group $C_k$, we specify the corresponding group representations as follows:
$L_g(s)$ rotates the observation image by $g \cdot a$ for action $a$, while
$R_g(a)$ rotates only the planar component $\mathbf{a}_{xy}$ by $g \cdot 90^\circ$ and leaves the auxiliary component $\mathbf{a}_{\mathrm{aux}}$ unchanged.

Actor network parametrized by $\xi$, critic network parametrized by $\phi$, and encoder are implemented using \texttt{E2CNN}~\cite{weiler2021e2cnns}, leveraging regular representations of $C_k$.
With this architecture, the policy $\pi_\xi : \mathcal{S} \rightarrow \mathcal{A}$ is explicitly constrained to be equivariant,
\begin{equation}
    \pi_{\xi}(L_g (s)) = R_g (\pi_{\xi}(s)), \quad \forall g \in C_4,
\end{equation}
which ensures that rotating the workspace observation results in a coherently rotated planar motion prediction.
In contrast, the critic is designed to be invariant to the joint transformation of state and action, yielding consistent value estimates across symmetric configurations:
\begin{equation}
    Q_\phi(e(L_g (s)), R_g (a)) = Q_\phi(e(s), a), \quad \forall g \in C_4,
\end{equation}
where $e(\cdot)$ denotes the equivariant feature encoder, and $Q_{\phi}$ is the $Q$-value function of the critic network $\phi$.

\begin{figure*}[!t]
    \centering
    \includegraphics[width=\textwidth]{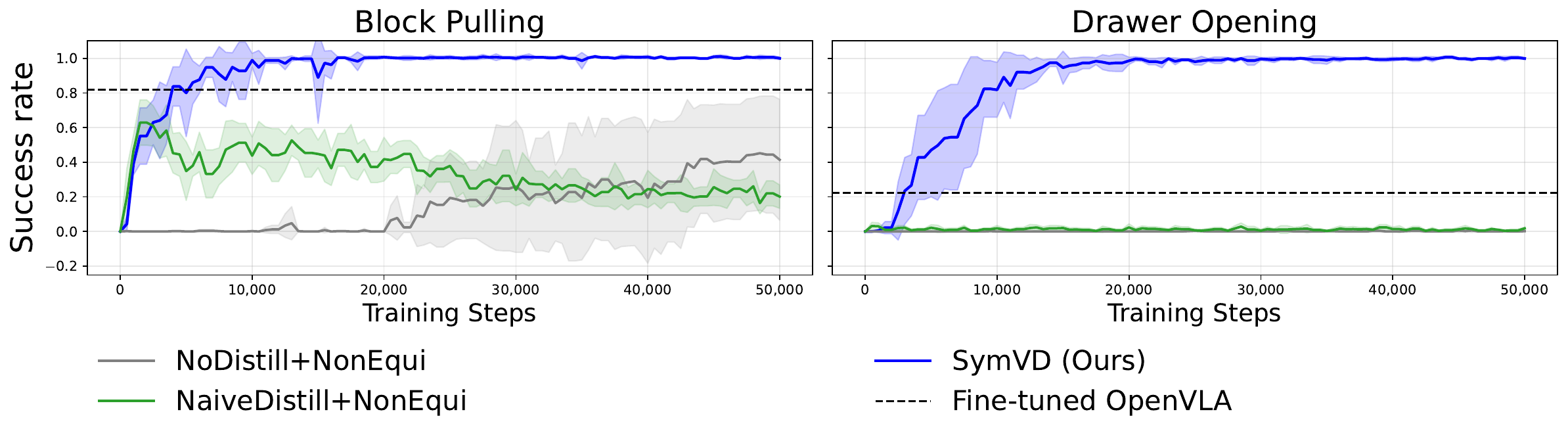}
    \caption{Comparison of evaluation rewards for OpenVLA teacher with various distillation methods on the Drawer Opening and Block Pulling tasks. 
    The dashed horizontal line denotes the performance of the fine-tuned OpenVLA. We compare student policies trained via naïve distillation and SymVD. Results are averaged over 10 random seeds.}
    \label{fig:results2}
\end{figure*}

\subsection{Group-Averaged Distillation}

We incorporate an additional distillation term to transfer knowledge from a pretrained VLA teacher $\pi_{\text{VLA}}$.
However, naïve pointwise distillation loss such as mean-squared error~\cite{julg2025refined} can be misaligned with equivariant policy architectures. The teacher may contain symmetry-breaking components, while the equivariant student is structurally constrained to be consistent under group transformations.

Recall the action decomposition $\mathcal{A}=\mathcal{A}_{\mathrm{eq}}\times \mathcal{A}_{\mathrm{inv}}$,
where the group action $R_g$ rotates the planar component $\mathbf a_{xy}\in\mathcal{A}_{\mathrm{eq}}$ and leaves
the auxiliary component $\mathbf a_{\mathrm{aux}}\in\mathcal{A}_{\mathrm{inv}}$ unchanged.
We define the group-averaging operator
\begin{equation}
P(a) := \mathbb{E}_{g\sim \mathrm{Unif}(G)}\,[R_g(a)].
\label{eq:action_averaging_main}
\end{equation}

\begin{lemma}[Orthogonal decomposition induced by action-space group averaging]
\label{lem:orthogonal_decomposition_main}
Assume $G$ is a finite group acting on the action space $\mathcal{A}$ through linear maps
$\{R_g:\mathcal{A}\to\mathcal{A}\}_{g\in G}$, and endow $\mathcal{A}$ with a $G$-invariant inner product
$\langle a,b\rangle := a^\top b$.\footnote{This holds, for instance, when $R_g$ is orthogonal for all $g\in G$.}
Then $P$ as defined in~\eqref{eq:action_averaging_main} is self-adjoint, idempotent (hence an orthogonal projector), and admits
\begin{equation}
\mathcal{A} = \mathrm{range}(P)\ \oplus\ \ker(P),
\qquad
a = P(a) + (I-P)a,
\label{eq:action_decomposition_main}
\end{equation}
with $\mathrm{range}(P)=\{a\in\mathcal{A}: R_g(a)=a,\ \forall g\in G\}$.
\end{lemma} 

A complete proof of~\eqref{eq:action_decomposition_main} can be found in~\cite{elesedy2021provablystrictgeneralisationbenefit};
we directly apply their argument to our action-space.
A consequence of Lemma~\ref{lem:orthogonal_decomposition_main} is that under our factorization $\mathcal{A}=\mathcal{A}_{\mathrm{eq}}\times \mathcal{A}_{\mathrm{inv}}$, where
$R_g$ rotates only $\mathbf a_{xy}\in\mathcal{A}_{\mathrm{eq}}$ and leaves $\mathbf a_{\mathrm{aux}}\in\mathcal{A}_{\mathrm{inv}}$
unchanged, we have $\mathrm{range}(P)=\mathcal{A}_{\mathrm{inv}}$ and $\ker(P)=\mathcal{A}_{\mathrm{eq}}$.
In particular, for symmetric rotation groups such as $G=C_k$, $P$ preserves $\mathbf a_{\mathrm{aux}}$ while canceling the directional component $\mathbf a_{xy}$ in expectation.

In our setting, the teacher observes a fixed front-view image, which does not admit the same state transformation $L_g$. We therefore apply group averaging directly in the action space and define the distillation objective as
\begin{equation}
\mathcal{L}_{\text{dist}}(\xi)
:=
\left\|
\mathbb{E}_{g\sim \mathrm{Unif}(G)}
\left[
R_g\!\left(\pi_\xi(s) - \pi_{\mathrm{VLA}}(s)\right)
\right]
\right\|^2.
\label{eq:dist_expect_main}
\end{equation} 
In practice, we compute~\eqref{eq:dist_expect_main} with invariant auxiliary
mask,
\begin{equation}
\mathcal{L}_{\mathrm{dist}}(\xi)
=
\left\|
\mathbf m \odot \left(\pi_\xi(s) - \pi_{\mathrm{VLA}}(s)\right)
\right\|^2
\label{eq:distill_main}
\end{equation}
where $\odot$ denotes element-wise multiplication and the binary mask $\mathbf m\in\{0,1\}^d$ selects only the $G$-invariant action components:
$m_i=\mathbbm{1}\{i\in\mathcal{I}_{\mathrm{aux}}\}$ with
$\mathcal{I}_{\mathrm{aux}}:=\left\{i\in\{1,\dots,d\}:\ \left(\mathbb{E}_{g\sim\mathrm{Unif}(G)}[R_g(a)]\right)_i=a_i,\ \forall a\in\mathcal{A}\right\}$.

Since $R_g$ leaves $\mathbf a_{\mathrm{aux}}$ unchanged,~\eqref{eq:distill_main} provides direct supervision on $\mathbf a_{\mathrm{aux}}$.
Moreover, by Lemma~\ref{lem:orthogonal_decomposition_main}, for symmetric rotation groups (e.g., $C_k$), the planar term is minimized at $\pi_{\xi,xy}(s)=\mathbf 0$ in expectation, so the planar component is left to be learned by reinforcement learning rather than being forced to match the teacher.

\paragraph{Justification under symmetric MDPs.}
When the MDP is perfectly $G$-invariant, there exists an optimal $G$-equivariant policy, which justifies restricting the student to an equivariant policy class.

\begin{lemma}[Existence of an optimal $G$-equivariant policy under $G$-invariant MDPs]
\label{lem:equiv_opt_policy_main}
Assume the MDP is $G$-invariant in the sense of ~\eqref{eq:group-invariant}, including $G$-invariance of the initial distribution $\rho_0$ and the terminal condition under $L_g$.
Then there exists an optimal policy $\pi_{\xi}^\star$ such that
\begin{equation}
\pi_{\xi}^\star(L_g (s)) = R_g \big(\pi_{\xi}^\star(s)\big), \qquad \forall g \in G.
\end{equation}
\end{lemma}

\noindent
A complete proof can be found in~\cite{wang2022mathrm}; Lemma~\ref{lem:equiv_opt_policy_main} guarantees the existence of an optimal equivariant policy under a G-invariant MDP. We apply the same result to our setting by instantiating the group actions as $(L_g, R_g)$, satisfying the invariance condition~\eqref{eq:group-invariant}. As a result, a student equipped with an equivariant architecture is inherently restricted to symmetry-consistent policies, motivating a distillation objective whose supervisory signal lies within the representational capacity of the student. Eq.~\eqref{eq:dist_expect_main} therefore provides a valid and well-aligned training signal.

\paragraph{Overall objective with adaptive distillation.}
We combine the standard SAC actor loss with adaptive distillation:
\begin{equation}
\mathcal{L}_{\mathrm{actor}}(\xi)
=
\mathcal{L}_{\pi}(\xi)
+
\lambda_t \, \mathcal{L}_{\mathrm{dist}}(\xi),
\label{eq:overall}
\end{equation}
where $\mathcal{L}_{\pi}$ denotes the SAC actor loss.
Following~\cite{zhao2022adaptivereg}, we adapt $\lambda_t$ using a proportional--derivative (PD) controller:
\begin{equation}
\begin{aligned}
\Delta \lambda_t
&=
K_P \big( R_{\mathrm{target}} - R_{\mathrm{avg},t} \big)
\\ &\quad
+ K_D \max\!\Big(0,\,
R_{\mathrm{avg},t} - R_{\mathrm{current},t}\Big),
\\
\lambda_{t+1}
&=
\Pi_{[0,\lambda_{\max}]}\!\Big(\lambda_t + \Delta \lambda_t\Big).
\end{aligned}
\label{eq:adaptive}
\end{equation}
We initialize $\lambda_0=1$ and gradually reduce $\lambda_t$ as the student approaches the target performance.

\begin{figure*}[!t]
    \centering
    \includegraphics[width=\textwidth]{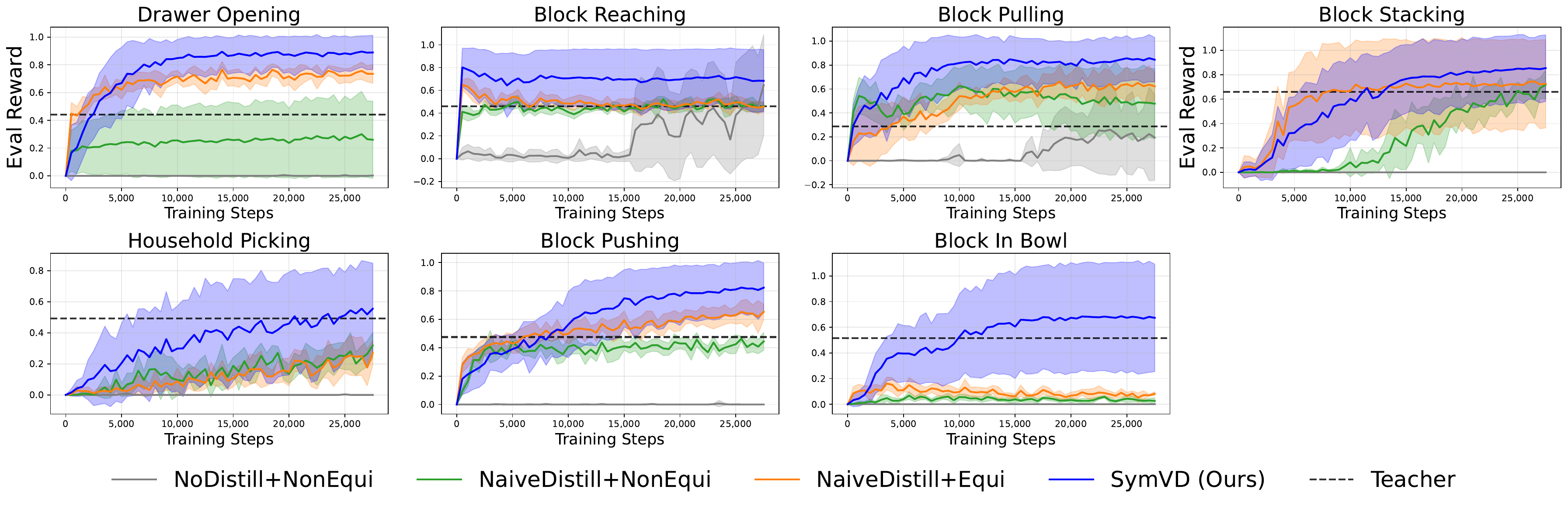}
    \caption{Comparison of evaluation rewards during training for SAC and Equivariant SAC under different distillation strategies across 7 tasks. We report results for no distillation, naïve distillation, and SymVD(ours), applied to both SAC and Equi-SAC student policies, all of which distill actions from a DrQ or TD3 expert. Performance is measured every 500 training steps and averaged over 10 different seeds. The dashed line denotes the pretrained teacher’s averaged reward over 100 episodes prior to distillation. Shaded regions indicate the standard error across runs.}
\label{fig:results}
\end{figure*}

\section{Experiment}
\label{sec:experiments}
\subsection{Task and Setup}
We conduct experiments on 7 closed-loop manipulation tasks in the BulletArm environment, with a 7-DoF robotic arm including a gripper holding state. Cartesian translations $(x, y, z)$ in action are bounded within $[-0.05, 0.05]$, and rotations around the $z$-axis are limited to $[-45^{\circ}, 45^{\circ}]$. The robot manipulation tasks include block pulling, drawer opening, block stacking, block reaching, block pushing, block-in-bowl, and object picking. In our experimental setting, a teacher policy is first trained (or fine-tuned), and subsequently used to provide action supervision for an equivariant student agent. We set $R_{\mathrm{target}} = \alpha \overline{R}_{\mathrm{teacher}}$ with $\alpha = 1.05$ in ~\eqref{eq:adaptive},
where $\overline{R}_{\mathrm{teacher}}$ is the average reward of the pretrained teacher policy.

\subsection{Baseline Experiment}\label{subsec:baseline}
We first evaluate the proposed distillation framework using a foundation model. For the teacher, we fine-tune OpenVLA-7B using LoRA, the apply the proposed distillation framework to transfer knowledge to the equivariant SAC student. Detailed training configurations are provided in Appendix~\ref{app:training_details}.

\paragraph{Experimental Configurations.}
We evaluate distillation under three training configurations:
\begin{enumerate}[label=(\arabic*)]
    \item NoDistill$+$NonEqui: A standard SAC agent trained from environment rewards, using a non-equivariant actor-critic architecture.
    \item Na\"iveDistill$+$NonEqui: The same non-equivariant SAC agent, augmented with a conventional pointwise distillation loss that directly matches the teacher's actions.
    \item SymVD (ours): An equivariant SAC agent, augmented with a symmetry-aware distillation objective that respects group transformations, together with an adaptive weighting scheme to balance distillation and RL updates.    
\end{enumerate}
Comparing (1) against (2) and (3) measures the benefit of adding teacher supervision. Comparing (2) versus (3) isolates the advantage of SymVD over conventional distillation by highlighting the effect of symmetry-aware distillation combined with an equivariant architecture. Note that the fine-tuned OpenVLA results are included as a teacher-performance reference.

\paragraph{Experimental Results.}As shown in Fig.~\ref{fig:results2}, SymVD consistently outperforms naïve distillation on both tasks, achieving higher evaluation rewards. While both distillation variants outperform non-distill SAC, naïve distillation saturates without surpassing the teacher.
performance. In contrast, SymVD enables the equivariant student to continue improving and surpass the fine-tuned VLA, indicating more effective knowledge transfer. Both no-distillation and naïve distillation in the Drawer Opening task fail to make meaningful learning progress, due to ineffective exploration in challenging tasks that require broad exploration under low-performing teacher. By removing unnecessary direction- and pose-specific action components, SymVD allows RL to focus on task refinement, leading to sustained performance improvements.

\subsection{Distillation with Lightweight Teachers}
To assess the generality of the proposed distillation framework, we conduct additional experiments using lightweight teacher policies across a broader set of manipulation tasks.
We consider teacher networks that are less powerful and less performative compared to the VLA-based teacher from Section.~\ref{subsec:baseline} to show that distilling from bad, low-quality teachers with SymVD still executes the task reasonably well.
As lightweight teachers, we employ RAD-DrQ with $\mathrm{SO}(2)$ data augmentation and TD3+BC. To reflect the practical regime of fine-tuned VLA models, which often exhibit limited success rates despite nontrivial task knowledge, teacher training is intentionally terminated once the averaged reward reaches 0.4. The averaged rewards of all teacher policies used in our experiments are reported in Appendix~\ref{tab:teacher_performance}. Before student learning, 100 demonstration episodes collected from the teacher are preloaded into the replay buffer.
\paragraph{Experimental Configurations.}
In addition to the baselines introduced in Section~\ref{subsec:baseline}, we include one additional configuration:
\begin{enumerate}[label=(\arabic*)]
\setcounter{enumi}{3}
    \item Na\"iveDistill$+$Equi: Equivariant SAC trained with the same na\"ive distillation loss as (2).
\end{enumerate}
All other settings are kept identical to those in Section~\ref{subsec:baseline}.
Comparing (2) and (4) isolates the effect of imposing equivariant architectural constraints under the same na\"ive distillation objective, while comparing (4) and (3) evaluates whether our proposed SymVD better aligns teacher supervision with the representational constraints of an equivariant student.

\paragraph{Experimental Results}The results on seven robotic manipulation tasks are summarized in Fig.~\ref{fig:results}. Across all tasks, our method demonstrates faster convergence and achieves the highest evaluation reward compared to all baselines.

The results further indicate that the role of distillation varies with task difficulty. For relatively low-difficulty tasks, such as Block Pulling and Block Reaching, the teacher facilitates early-stage exploration, allowing the student to refine its behavior with distillation. In these settings, even naïve distillation achieves moderate performance improvements through relying on suboptimal teacher supervision. In contrast, for high-difficulty tasks requiring precise alignment, such as Object Picking and Block In Bowl, the role of distillation shifts from accelerating initial exploration to shaping the overall inductive bias of the policy. Here, naïve distillation—regardless of architectural equivariance—tends to preserve pose- and direction-specific biases from the teacher, which interfere with consistent policy refinement throughout training. SymVD addresses this challenge by removing such symmetry-breaking biases from teacher supervision and aligning the student with the teacher only in symmetry-consistent subspaces. 

In the block pulling and picking tasks, equivariant SAC trained with naïve distillation sometimes exhibits comparable or even inferior performance to SAC under the same objective, despite its inherent advantage in exploiting transformation equivariance. This degradation highlights a mismatch between symmetry-agnostic distillation objectives and equivariant policy, as conventional distillation disregards the structured inductive biases encoded in the network.

\subsection{Adaptive vs. Non-Adaptive Distillation}
\begin{figure}[t]
    \centering
    \includegraphics[width=\columnwidth]{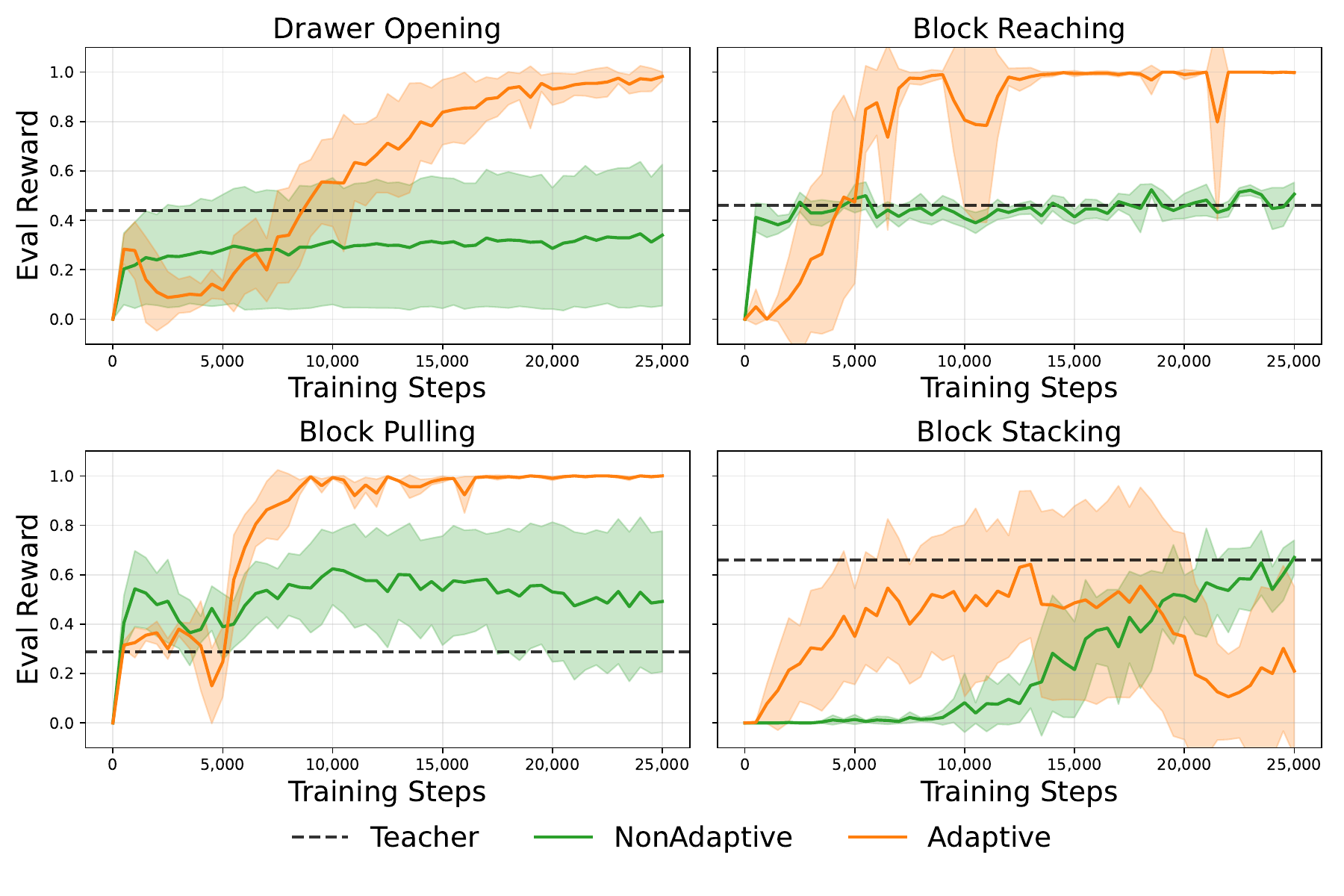}
    \caption{The effect of adaptive $\lambda$ in naïve distillation applied to SAC. 
    We compare evaluation rewards between an adaptive distillation weight updated according to ~\eqref{eq:adaptive} and a non-adaptive baseline with a constant $\lambda = 1$. 
    Performance is evaluated every 500 training steps and averaged across multiple random seeds. 
    Shaded regions indicate the standard error across runs.}
    \label{fig:adaptive_results}
\end{figure}

We investigate the effect of adaptive distillation weighting by comparing a fixed distillation weight (\(\lambda = 1\)) in~\eqref{eq:overall} with the proposed adaptive \(\lambda\) schedule in~\eqref{eq:adaptive}. To isolate the impact of the weighting strategy, all experiments are conducted using naïve SAC without equivariant architectures. The results are shown in Fig.~\ref{fig:adaptive_results}. Across most tasks, the adaptive scheme achieves higher evaluation rewards than the non-adaptive baseline, indicating that a constant distillation weight can cause performance to plateau when the teacher policy is suboptimal. For the \emph{Block Stacking} task, the adaptive scheme improves performance in the early stages but exhibits performance degradation during later training. Since distillation from a weak teacher with a 40\% success rate fails to converge in this setting, we instead employ a stronger teacher with a success rate exceeding 62\%. These results suggest that while distillation is effective for guiding early learning, excessive reliance on teacher can hinder stable policy improvement in complex tasks.

\section{Conclusion}
\label{sec:conclusion}
In this work, we introduced Symmetric VLA Distillation (SymVD), a framework designed to transfer knowledge from pretrained VLA teachers to compact reinforcement learning policies by leveraging geometric symmetries in robotic manipulation.
SymVD integrates an equivariant actor-critic architecture with a symmetry-aware distillation objective. This alignment ensures that teacher supervision respects the student's representational constraints, thereby effectively reducing redundant exploration across symmetry-related configurations.
Furthermore, we proposed an adaptive weighting scheme that dynamically balances distillation and reinforcement learning updates, enhancing training stability and mitigating performance stagnation when teacher signals are imperfect.

Empirically, our experiments on BulletArm manipulation tasks demonstrate that SymVD consistently outperforms no-distilled SAC and conventional distillation in terms of sample efficiency and generalization.
We further highlighted the robustness of SymVD when distilling from lightweight and low-quality teachers, confirming that symmetry-consistent supervision together with adaptive weighting is essential for avoiding over-reliance on suboptimal demonstrations.

In conclusion, our results suggest that incorporating task symmetries into the distillation pipeline is a practical and effective strategy for compressing generalist VLA knowledge into deployable policies.
Future work will explore extending SymVD to richer symmetry groups and diverse embodiments, as well as integrating state-level transformations to facilitate fully symmetry-aware distillation beyond the action space.

\clearpage
\bibliographystyle{named}
\bibliography{ijcai26}

\newpage
\appendix
\onecolumn

\section{Vision-Language-Action Fine-Tuning}
\label{app:training_details}
\begin{figure}
    \centering
    \subfloat[Front-view]{%
    \includegraphics[width=0.2\linewidth]{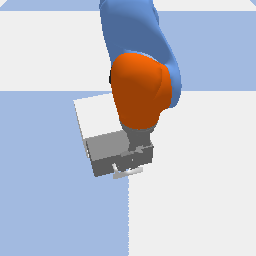}
    }
    \subfloat[Eye-in-hand]{%
    \includegraphics[width=0.2\linewidth]{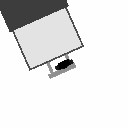}
    }
    \caption{Synchronized observation setup for OpenVLA-based distillation.}
    \label{fig:obs_setup}
\end{figure}

We fine-tune OpenVLA-7B using LoRA, following the officially released \texttt{OpenVLA-OFT} recipe. 
As PyBullet environments do not provide language instructions, we convert each task name (e.g., \texttt{CloseLoopDrawerOpening}) into a short textual description specifying the manipulation task, which is used only in the VLA-based extension. (see Table~\ref{tab:task_instructions}). This text is used solely for the VLA-based teacher and is tokenized together with the corresponding action sequence. 
The model is trained for up to 200K steps with a learning rate of $5\times10^{-4}$. 
We adopt an $\ell_1$ regression objective to predict multi-step action chunks of size $4$, which are decoded in parallel. 
LoRA modules with rank $32$ are applied to all linear layers, resulting in 1.45\% trainable parameter relative to the frozen base model. 
The LoRA weights are initialized using a Gaussian distribution. During training, the replay buffer stores up to 100{,}000 transitions. The detailed hyperparameters are summarized in Table~\ref{tab:training_details}.
Since OpenVLA requires third-person front-view images that capture the robot arm but do not support environmental transformations, we use only the front-view observations to extract teacher actions. In contrast, the RL agent receives the eye-in-hand view as input, as illustrated in Fig.~\ref{fig:obs_setup}.

\begin{table}[t]
\centering
\begin{tabular}{lp{9cm}}
\hline
Task name & Instruction \\
\hline
\texttt{close\_loop\_block\_pulling} & Pull one of the two blocks to make contact with the other block. \\
\texttt{close\_loop\_drawer\_opening} & Pull the handle to open the drawer. \\
\hline
\end{tabular}
\caption{Textual task instructions used for VLA fine-tuning.}
\label{tab:task_instructions}
\end{table}

\begin{table}[t]
\centering
\begin{center}
\begin{tabular}{ p{4cm}||p{4cm} }
 \hline
 \textbf{Hyperparameter} & \textbf{Value} \\
 \hline
 \#GPUs & H100 \\
 Optimizer / Learning rate & Adam ($5\times10^{-4}$) \\
 Batch size & 16 \\
 Maximum training steps & 200,000 \\ 
 Steps before decay & 10,000 \\
 Validation frequency & 200,000 \\
\hline
 \#Training Configs & \\ 
 Shuffle buffer size & 100,000 \\
 Use L1 regression & True \\
 Use FiLM & False \\
 Use proprioception & False \\
 Number of input images & 1 \\
 \hline
 \#Input and Augmentation & \\
 Image size & $224 \times 224$ px \\
 Image augmentations & Yes \\
 Action normalization & Yes \\
 \hline
 \#LoRA Configs & \\
 Action chunk size & 4 \\
 LoRA rank & 32 \\
 LoRA dropout & 0 \\
\hline
 \#Trainable parameters & 110,828,288(1.448\%) of 7,652,065,472 total \\
 \hline
\end{tabular}
\end{center}
\caption{Table of hyperparameters for OpenVLA-7B fine-tuning on BulletArm.}
\label{tab:training_details}
\end{table}
\section{Training for Lightweight Teacher Policies}
We employ pretrained teacher policies based on either DrQ-fD or TD3+BC. DrQ is an image-based off-policy RL algorithm that applies data augmentation to pixel observations, (e.g., random shifts) and trains the critic and actor using the augmented observations. In this work, we adopt DrQ from Demonstrations (DrQ-fD), which optimizes DrQ with an additional mean squared error loss between the agent policy and expert demonstration. TD3+BC combines Twin Delayed DDPG (TD3) with a behavior cloning objective to constrain the learned policy toward the demonstration data. Additionally, to facilitate stable training of the lightweight teacher policies, we prefill the replay buffer with more than 500 demonstration episodes generated by an environment-based planner. For \texttt{drawer\_opening}, \texttt{block\_reaching}, and \texttt{block\_pulling}, we use DrQ teachers trained with random-shift augmentation. For the remaining tasks, since DrQ training did not reliably succeed, we instead employ TD3+BC teachers. The averaged rewards of all teacher policies used in our experiments are reported in Table~\ref{tab:teacher_performance}.
For comparison, we implement a naïve distillation objective following prior work~\cite{julg2025refined} and apply it to the SAC actor.
\begin{equation}
\mathcal{L}_{\mathrm{dist}}^{\mathrm{naive}}(\xi)
=\left\|
\pi_\xi(s_t) - \pi_{\mathrm{VLA}}(s_t)
\right\|^2,
\label{eq:distill_naive}
\end{equation}
where $\pi_\xi(s_t)$ denotes the mean action predicted by the student SAC actor and $\pi_{\mathrm{VLA}}(s_t)$ is the teacher action produced by the VLA policy.
\begin{table}[t]
    \centering
    \begin{tabular}{lcc}
        \hline
        Task & Teacher Algorithm & Average reward \\
        \hline
        Block Pulling   & DrQ-fD   & 0.287 \\
        Drawer Opening  & DrQ-fD   & 0.441 \\
        Block Reaching  & DrQ-fD   & 0.460 \\
        Block Stacking  & TD3+BC  & 0.660 \\
        Object Picking  & TD3+BC   & 0.493 \\
        Block Pushing   & TD3+BC   & 0.475 \\
        Block in Bowl   & TD3+BC   & 0.514 \\
        \hline
    \end{tabular}
    \caption{Performance of the pretrained teacher policies used in the baseline experiments, evaluated over 100 episodes.}
    \label{tab:teacher_performance}
\end{table}
\end{document}